\documentclass[conference]{IEEEtran}
\IEEEoverridecommandlockouts
\usepackage{cite}
\usepackage{amsmath,amssymb,amsfonts}
\usepackage{algorithmic}
\usepackage{graphicx}
\usepackage{textcomp}
\usepackage{xcolor}
\usepackage{amsmath,bm}
\usepackage{subcaption}

\usepackage{algorithm}
\usepackage{algorithmic}

\usepackage[letterpaper, top=0.767in, bottom=0.9in, left=0.68in, right=0.693in, includeheadfoot]{geometry}

\begin{document} 

\title{SCDM: Score-Based Channel Denoising Model for Digital Semantic Communications\\

}

\author{
    \IEEEauthorblockN{Hao Mo$^{1,2}$, Yaping Sun$^{1,2}$, Shumin Yao$^{1}$, Hao Chen$^1$, Zhiyong Chen$^4$, Xiaodong Xu$^{5,1}$,}
    \IEEEauthorblockN{Nan Ma$^{5,1}$, Meixia Tao$^{4}$, Shuguang Cui$^{2,3,1}$}
    \IEEEauthorblockA{$^1$ Department of Broadband Communication, Pengcheng Laboratory, Shenzhen 518066, China}
    \IEEEauthorblockA{$^2$ FNii, the Chinese University of Hong Kong (Shenzhen), Shenzhen 518172, China}
    \IEEEauthorblockA{$^3$ SSE, the Chinese University of Hong Kong (Shenzhen), Shenzhen 518172, China}
    \IEEEauthorblockA{$^4$ CMIC, Shanghai Jiao Tong University, Shanghai 200240, China}
    \IEEEauthorblockA{$^5$ Beijing University of Posts and Telecommunications, Beijing 100876, China}
    \IEEEauthorblockA{mohao.ai@outlook.com, \{sunyp,yaoshm,chenh03,xuxd\}@pcl.ac.cn}
    \IEEEauthorblockA{manan@bupt.edu.cn, \{zhiyongchen, mxtao\}@sjtu.edu.cn, shuguangcui@cuhk.edu.cn}
}

\maketitle

\begin{abstract}
    Score-based diffusion models represent a significant variant within the family of diffusion 
    models and have found extensive application in the increasingly popular domain of generative tasks.
    Recent investigations have explored the denoising potential of diffusion models in 
    semantic communications. However, in previous paradigms, noise distortion in the diffusion process 
    does not match precisely with digital channel noise characteristics. In this work, we introduce the Score-Based Channel Denoising Model (SCDM) for 
    Digital Semantic Communications (DSC). SCDM views the distortion of constellation symbol sequences in digital transmission as a score-based forward diffusion process. 
    We design a tailored forward noise corruption to better align digital channel noise properties in the training phase. During the inference stage, the well-trained SCDM can 
    effectively denoise received semantic symbols under various SNR conditions, reducing the difficulty for the semantic decoder in extracting semantic information 
    from the received noisy symbols and thereby enhancing the robustness of the reconstructed semantic information. Experimental results show that SCDM outperforms the 
    baseline model in PSNR, SSIM, and MSE metrics, particularly at low SNR levels. Moreover, SCDM reduces storage requirements by a factor of 7.8. 
    This efficiency in storage, combined with its robust denoising capability, makes SCDM a practical solution for DSC across diverse channel conditions.
\end{abstract} 
  
\section{Introduction}
In the realm of the sixth generation (6G) mobile communication network development, a profound integration with artificial intelligence is imperative. Semantic communications, 
 an innovative form of communication, leverages deep learning to 
design joint source-channel coding. This approach effectively addresses the intelligent task requirements 
within 6G networks. It not only further reduces transmission overhead but also ensures robust performance 
under low signal-to-noise ratios (SNR) \cite{10634888} \cite{10615729}.   
\begin{figure}[ht] 
    \centering
    \includegraphics[width=1.0\linewidth]{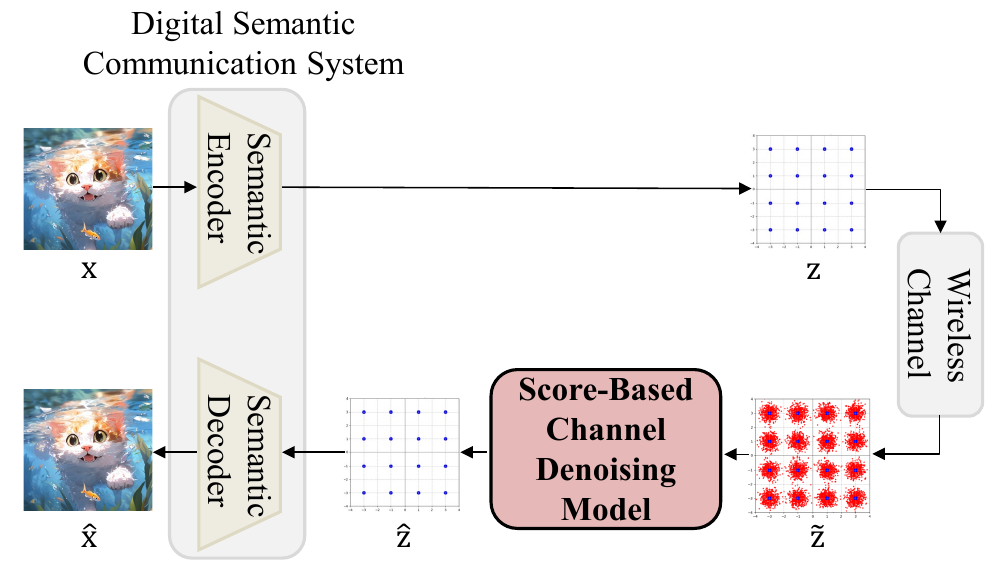}  
    \caption{An essential structure of our proposed SCDM-DSC.}
    \label{fig:1}  
\end{figure} 

Early works on semantic communication mainly focus on analog transmission, where the continuous-valued feature vectors generated by neural network-based encoders are transmitted 
directly in an analog fashion. These schemes have demonstrated significant potential and are often referred to as Analog Semantic Communications (ASC). For instance, 
Deep Joint Source-Channel Coding (Deep-JSCC) \cite{8683463} integrates deep learning with JSCC and has outperformed conventional digital transmission methods, 
including those using JPEG compression. To further enhance the semantic reconstruction capabilities of JSCC, the researchers in \cite{10480348} introduce a Channel Denoising Diffusion Model (CDDM) which utilizes diffusion model for wireless channel 
denoising, facilitating noise reduction of analog semantic information under varying channel SNR 
conditions.     
 
 
However, ASC is difficult to implement in practice due to hardware limitations, such as power amplifier imperfections and finite-resolution analog-to-digital converters. Thus, Digital Semantic Communications (DSC), dedicated to 
nowadays widespread digital communication systems today, has emerged as a promising alternative. DSC discretizes semantic information for transmission. The three fundamental approaches are as follows.
 The first simply converts each element in the continuous-valued feature vectors into bits with 
 full-precision and transmits them using digital modulation \cite{xie2021lite,10287247}. 
 The second approach leverages vector quantization techniques, such as Vector Quantized Variational Autoencoder (VQ-VAE) \cite{oord2017neural} 
 or Vector Quantized Generative Adversarial Network (VQ-GAN) \cite{esser2021taming}, to discretize the feature space, necessitating a shared 
 knowledge base between the transmitter and receiver \cite{zhou2024moc}. Unlike the previous two approaches where the discretization is performed 
 separately from semantic coding, the third approach is a joint coding and modulation (JCM) framework \cite{bo2024joint}, which utilizes variational autoencoder (VAE) 
 and probabilistic sampling tool (such as Gumbel softmax) to learn and generate the discretized feature symbols directly. 
  
Although these methods in ASC/DSC have shown promising results, they are still suffering from the following challenges:
\begin{itemize}
    \item \textbf{Susceptibility to high SNR variance}.  
    The JSCC training scheme implicitly incorporates both channel noise reduction and error correction for data reconstruction 
    within the codec. As a result, the neural networks of the codec need to adapt to the semantic symbols corrupted by noise over a wide range 
    of SNR levels, presenting significant challenges to the training of the codec.
    \item \textbf{Substantial storage requirements}. Certain methods require 
    training distinct codec models for each discrete SNR level, leading to increased storage consumption, which may impede practical implementation, 
    especially on mobile devices.
    \item \textbf{Mismatch with digital communication systems}. Although CDDM exhibits robust noise-resistance capabilities and shows promise for 
    handling a wide range of SNR levels within a single model, the inherent noise distortion in its diffusion process does not precisely align 
    with the noise characteristics of digital channels due to the \textit{drift} term of the forward diffusion not being zero. 
    This limitation prevents its direct application in DSC.
\end{itemize}
    
To address these challenges, we propose a Score-Based Channel Denoising Model (SCDM) module specifically designed for DSC. 
First, SCDM utilizes an independent score-based diffusion model to denoise the received digital semantic symbol before decoding. 
SCDM iteratively denoises the corrupted semantic symbol, producing a restored symbol with low variance relative to the semantic symbol originally sent by the transmitter. 
This independence enables SCDM to function as a plug-and-play module without requiring integration into an existing JSCC training scheme. 
Second, our model can handle a broad range of SNR levels 
within a single architecture, significantly reducing the storage resource requirements. 
Third, we align the noise distortion process in SCDM’s diffusion with the inherent noise characteristics of digital channels, 
enhancing its suitability for digital semantic symbol denoising.

In sum, our work makes the following contributions:
\begin{itemize}
\item We propose a novel SCDM module for constellation symbol denoising in DSC. The training process of 
score-based diffusion model is closely aligned with the noise distortion process of digital channels. The trained score-based diffusion model has learned 
the noise distortion process of the digital communication channel, and thereby SCDM can effectively denoise the digital semantic symbol under various SNR conditions. 

\item We introduce a unified model architecture capable of handling a broad range of SNR levels. 
This reduces the need for multiple codec pairs, thus significantly lowering storage requirements. 
The unified approach not only enhances model efficiency but also simplifies deployment on resource-constrained mobile devices.

\item Our simulations show that our model consistently outperforms baseline method JCM (single codec) in both PSNR and SSIM across a wide range of SNR levels, 
with significant MSE improvements in low SNR conditions, enhancing system robustness and easing the adaptation of the semantic decoder to varying noise conditions.
\end{itemize}

\section{System Model}

In this scection, we introduce the system model of the proposed DSC framework with SCDM module. We term it as SCDM-DSC for expression convenience. 
As illustrated in Fig. \ref{fig:1}, the framework consists of two primary components: a typical DSC system featuring a semantic encoder at the transmitter and a decoder at the receiver \cite{bo2024joint,zhou2024moc}, and our proposed SCDM module. 
The details of each component are described as follows.

At the transmitter, a deep-neural-network-based (DNN-based) semantic encoder $\mathcal{E}$ parameterized by $\phi$ takes a source message $\mathbf{x} \in \mathcal{X}$, 
where $\mathcal{X}$ represents the dataset (e.g., a collection of images), as input and transforms it into a sequence of constellation 
symbol \(\mathbf{z}\) for transmission. 
Here, \(\mathbf{z} \in \mathbb{C}^n\) and \(n\) is the number of channel uses. The elements of \(\mathbf{z}\) 
are drawn from a set \(\mathcal{Z} = \{ z_1, z_2, \ldots, z_M \}\), where \(M\) represents the order of digital modulation.

We norm the symbol sequence $\mathbf{z}$ with average transmit power constraint $\mathit{P}$ before transmitting,
where $P = \frac{\|\mathbf{z}\|^2}{n}$. $\mathbf{z}$ is then transmitted to the receiver through a additive white Gaussian channel with 
noise. During this transmission, non-ideal channel conditions (such as AWGN) modify $\mathbf{z}$, 
resulting in a distorted version, $\mathbf{\tilde{z}}$, at the receiver.
\begin{equation}
    \label{eq:-1}
    \mathbf{\tilde{z}} = \mathbf{z} + {\sigma}\mathbf{\bm{\varepsilon}}, 
\end{equation}
where 
$\mathbf{\bm{\varepsilon}} \sim \mathcal{CN}(\mathbf{0},\mathbf{I})$. 

The receiver next adopt our propesed SCDM to denoise $\mathbf{\tilde{z}}$. While performing denoising, the SCDM aims to minimize the distance between 
$\mathbf{\hat{z}}$ and $\mathbf{z}$. That is,
\begin{equation}
    \label{eq:0}
    \min_{\mathbf{\hat{z}}, \mathbf{z}} \| \mathbf{\hat{z}} - \mathbf{z} \|.
\end{equation}

Finally, the receiver employ a DNN-based semantic decoder $\mathcal{D}$ parameterized by $\varphi$ to recover $\mathbf{x}$ from the denoised symbol $\mathbf{\hat{z}}$, obtaining the reconstruction message $\mathbf{\hat{x}}$.

\section{Score-Based Channel Denoising Model} 

In this section, we begin with the preliminary of score-based generative model \cite{song2020score}. Then, 
we describe the two stages of our training process: the initial training process for our SCDM and the subsequent 
joint training process. In the first stage, we focus on training the SCDM to establish our semantic symbol denoising module. 
In the second stage, we jointly train the SCDM and DSC, optimizing the semantic decoder $\mathcal{D}$ of the DSC for further 
performance improvement. Note that the score-based diffusion model can be represented in both continuous and discrete forms. 
In the following sections, we use ``$i$'' to denote the discrete representation and ``$t$'' to denote the continuous representation.


\subsection{Preliminary}

Score-based generative models \cite{vincent2011connection, song2019generative} are a class of models that progressively 
corrupt training data with increasing noise and then use a deep neural network to reverse this process by learning the score---a vector 
filed pointing to the direction where the likelihood of the data increasing the fastest. They are generally governed by 
a \textbf{forward} stochastic differential equation (SDE) process and a \textbf{backward} SDE process \cite{song2020score}, 
both of which have continuous and discrete forms. 
The continuous forward SDE process can be wtitten as:
\begin{equation}
    \label{eq:1}
    d\mathbf{z} = \mathbf{f}(\mathbf{z}, t) dt + g(t) d\mathbf{w},
\end{equation}
where \(\mathbf{z}\) is sampled from a known dataset \(p(\mathbf{z})\), \(\mathbf{w}\) is a standard Wiener process, 
\(\mathbf{f}(\mathbf{z}, t)\) is the \textit{drift coefficient} of \(\mathbf{z}\), and \(t \in [0, T]\).

After defining the forward diffusion process, the corresponding reverse-time SDE can be derived as follows:
\begin{equation}
    \label{eq:2}
   d\mathbf{z} = \left[\mathbf{f}(\mathbf{z}, t) - g(t)^2 \nabla_{\mathbf{z}} \log p_t(\mathbf{z})\right] dt + g(t) d\mathbf{\bar{w}},
\end{equation}
where \(\bar{\mathbf{w}}\) is also a standard Wiener process, and $\nabla_{\mathbf{z}} \log p_t(\mathbf{z})$ is the score fuction. 
Score matching techniques \cite{hyvarinen2005estimation, vincent2011connection} are then applied to approximate the score function using the following optimal function:
\begin{equation}
    \label{eq:3}
    \begin{split}
       \theta^{*} = &\underset{\theta}{\arg\min} E_{t}\Bigg\{ \lambda(t) E_{\mathbf{z}_0} E_{\mathbf{z}_t \mid \mathbf{z}_0} \Bigg[ \left| s_{\theta}(\mathbf{z}_t, t) \right. \\
                    & \left. - \nabla_{\mathbf{z}_t} \log p_{0 t}(\mathbf{z}_t \mid \mathbf{z}_0) \right|_{2}^{2} \Bigg] \Bigg\}.
    \end{split}
\end{equation}
Here, \(\lambda: [0, T] \to \mathbb{R}_{>0}\) serves as a positive weighting function, and \(t\) is drawn uniformly from the 
interval \([0, T]\). Additionally, \(s_{\theta^{*}}(\mathbf{z}_t, t)\) is the optimal solution of \eqref{eq:3} and can be used 
to approximate \(\nabla_{\mathbf{z}} \log p_t(\mathbf{z})\), as long as sufficient data is provided.


\subsection{Stage 1: Training of SCDM}\label{AA}
\textbf{\textit{Forward process of SCDM}}. Consider an AWGN channel in digital communications. The received 
symbol $\mathbf{\tilde{z}}$ can be viewed as the result of an iterative Gaussian corruption process. 
Here, we set $\mathbf{z}_i = \mathbf{\tilde{z}}$, allowing us to decompose the received symbol $\mathbf{\tilde{z}}$ as follows:
\begin{equation}
    \label{eq:4}
    \begin{split}
    \mathbf{z}_i &= \mathbf{z}_{i-1} + \sqrt{\beta_i} \mathbf{\bm{\varepsilon}}^*_{i-1} \\
        &= \mathbf{z}_{i-2} + \sqrt{\beta_{i-1}} \mathbf{\bm{\varepsilon}}^*_{i-2} + \sqrt{\beta_i} \mathbf{\bm{\varepsilon}}^*_{i-1} \\
        &= \mathbf{z}_{i-2} + \sqrt{\beta_{i-1} + \beta_i} \mathbf{\bm{\varepsilon}}_{i-2} \\
        &\dots \\
        &= \mathbf{z}_0 + \sqrt{\sum_{k=1}^{i} \beta_k} \mathbf{\bm{\varepsilon}}_0 \\
        &= \mathbf{z}_0 + \sqrt{{\bar{\beta}}_i} \mathbf{\bm{\varepsilon}}_0, \\
    \end{split}
\end{equation}
where \(\{\mathbf{\varepsilon}_i^*, \mathbf{\varepsilon}_i\}_{i=0}^N \overset{\text{iid}}{\sim} \mathcal{CN}(0, \mathbf{I})\), 
and \(\bar{\beta}_i\) corresponds to the noise level.
\begin{figure}[htbp] 
    \centering
    \begin{subfigure}[b]{0.3\linewidth} 
        \centering
        \includegraphics[width=\linewidth]{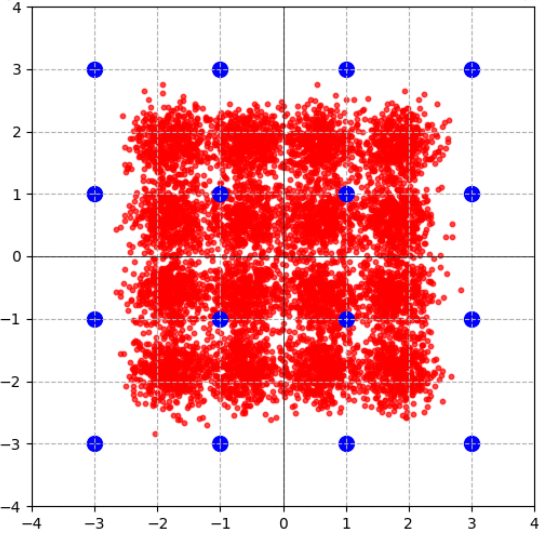} 
        \caption{CDDM at step $i$=16.}
        \label{fig:cddm_scdm_0} 
    \end{subfigure}
    \hfill
    \begin{subfigure}[b]{0.3\linewidth}
        \centering 
        \includegraphics[width=\linewidth]{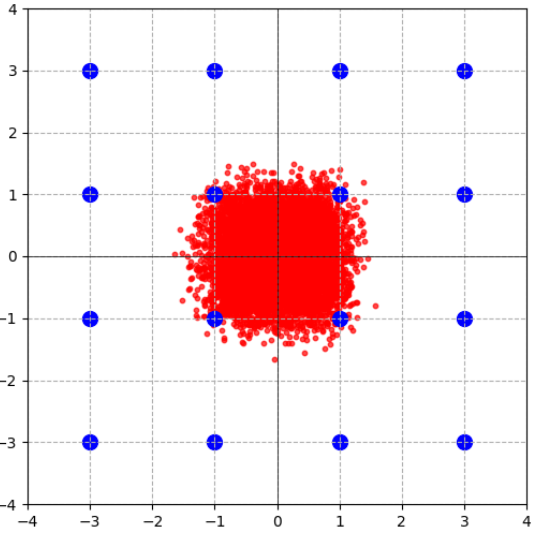} 
        \caption{CDDM at step $i$=64.}
        \label{fig:cddm_scdm_1}
    \end{subfigure}
    \hfill 
    \begin{subfigure}[b]{0.3\linewidth}
        \centering 
        \includegraphics[width=\linewidth]{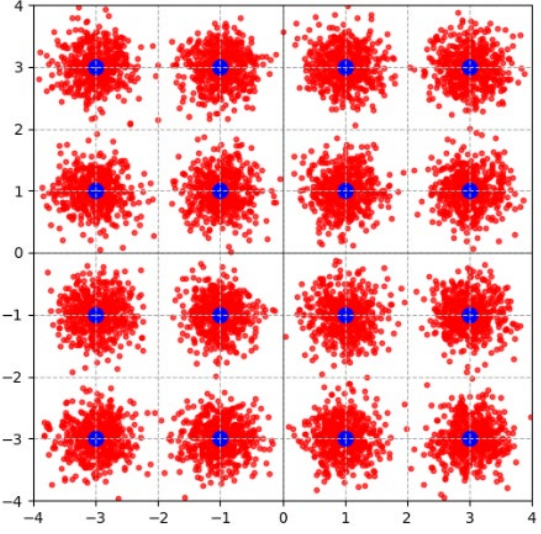}
        \caption{SCDM at step $i$=64.}
        \label{fig:cddm_scdm_2}
    \end{subfigure}
    \caption{\small Comparison of CDDM and SCDM of the forward diffusion process. The final diffusion
    step $N$ both set as 64. }
    \label{fig:cddm_scdm_comparison} 
\end{figure}
The training algorithm of the SCDM is summarized in Algorithm \ref{alg:scdm_training}. 
\begin{algorithm}[htbp]
    \caption{Training algorithm for SCDM}
    \label{alg:scdm_training}
    \begin{algorithmic}[1]
        \STATE \textbf{Input:} $\mathcal{X}$, $N$, $\sigma_i$, $\mathcal{E}$, $\phi$, $\theta$
        \STATE \textbf{repeat}

            \STATE  $\mathbf{x}$ $\sim$ $\mathcal{X}$, $i$ $\sim$ $\text{Uniform}(\{1, \ldots, N\})$
            \STATE $\mathbf{z}_{0}$ = $\mathcal{E}(\phi)$ ($\mathbf{x}$), $\mathbf{\bm{\varepsilon}}$ $\sim$ $\mathcal{CN}(0, \mathbf{I})$, compute $\mathbf{z}_{i}$ via \eqref{eq:5}
            \STATE Take gradient descent step on the discretize form of \eqref{eq:3}
            \[
            \nabla_\theta \left( \| s_{\theta}(\mathbf{z}_i, i) - \nabla_{\mathbf{z}_i} \log p_{0 i}(\mathbf{z}_i \mid \mathbf{z}_0)
            \|_2^2 \right)
            \]
        \STATE \textbf{until} Converged
    \end{algorithmic}
\end{algorithm}

We hereafter use \textit{annealed} diffusion forward process that align with the iteratively noise distortion process of an 
AWGN channel in digital communications as described in \eqref{eq:4} above. The discrete forward process of the SCDM is defined as:
\begin{equation}
    \label{eq:5}
    \mathbf{z}_i = \mathbf{z}_{i-1} + \sqrt{\sigma_i^2 - \sigma_{i-1}^2} \mathbf{\bm{\varepsilon}}, i \in \{1,\cdots,N\},
\end{equation}
where $N$ is a hyperparameter that determines the number of diffusion steps.
According to \cite{song2020score}, the corresponding continuous SDE expression of \eqref{eq:5} can be written as: 
\begin{equation}
\label{eq:6}
d\mathbf{z} = \sqrt{\frac{d\left[\sigma^2(t)\right]}{dt}} \, d\mathbf{w},
\end{equation}
where \(\sigma(t) = \sigma_{\text{min}} \left(\frac{\sigma_{\text{max}}}{\sigma_{\text{min}}}\right)^{t}\) for \( t \in [0,1] \).
The variance \(\sigma(t)\) increases gradually from \(\sigma_{\text{min}}\) to \(\sigma_{\text{max}}\), mimicking an annealing process, and is therefore referred to as an \textit{annealed} diffusion process. 
In equation \eqref{eq:6}, we set \(\mathbf{f}(\mathbf{z}, t) = 0\) and \(g(t) = \sqrt{\frac{d\left[\sigma^2(t)\right]}{dt}}\) in \eqref{eq:1}.
It is noteworthy that, unlike CDDM, we omit the \textit{drift coefficient} \(\mathbf{f}(\mathbf{z}, t)\) prior to \(\mathbf{z}_{t-1}\), 
as it would lead to the variance of the noisy constellation points \(\mathbf{z}_{t}\) collapsing around the origin of the constellation diagram, 
which is undesirable for digital communication systems in AWGN channels. We illustrate how the \textit{drift coefficient} affects the noise distortion in Fig. \ref{fig:cddm_scdm_comparison}.

\textbf{\textit{Backward process of SCDM}}. In the above the forward SDE process design, we have set the \textit{drift coefficient} $\mathbf{f}(\mathbf{z}, t)$ and the diffusion coefficient $g(t)$. 
According the reverse-time SDE \eqref{eq:2}, we can obtain the backward SDE of the SCDM as follows:
\begin{equation}
    \label{eq:7}
    d\mathbf{z} = -d\sigma^2(t) \nabla_{\mathbf{z}} \log p_t(\mathbf{z}) \, + \sqrt{\frac{d\sigma^2(t)}{dt}} \, d\mathbf{\bar{w}}
\end{equation}
According to \cite{song2020score},  $d\mathbf{\bar{w}} / \sqrt{dt} $ can be approximate as $\mathbf{\bm{\varepsilon}}(t) \sim \mathcal{CN}(\mathbf{0},\mathbf{I})$, when $dt \to 0$. Then we 
use \(s_{\theta^{*}}(\mathbf{z}_t, t)\) to substitute $\nabla_{\mathbf{z}} \log p_t(\mathbf{z})$, we can obtain the discrete backward process of the SCDM as follows:
\begin{equation}
    \mathbf{z}_i = \mathbf{z}_{i+1} + (\sigma_{i+1}^2 - \sigma_{i}^2)\mathbf{s_{\theta^{*}}}(\mathbf{z}_{i+1}, i+1) + \sqrt{\sigma_{i+1}^2 - \sigma_{i}^2} \mathbf{\bm{\varepsilon}}. 
\end{equation}
This discrete backward process provides a way to iteratively denoise the noise symbol $\mathbf{\tilde{z}}$ by solving the reverse-time SDE with numerical methods. The denoising steps \(N_{\text{SNR}}\) 
depend on the channel conditions, i.e., the noise levels, in our model. The numerical method we use is \textit{Predictor-Corrector} sampler of Variance Exploding (VE) SDE \cite{song2020score}, the \textit{Predictor}
part is the reverse diffusion SDE solver, and the \textit{Corrector} part is annealed Langevin dynamics \cite{song2019generative}.
The predictor part is described as \eqref{eq:7}. The Corrector applies Langevin dynamics to refine the Predictor term $\mathbf{z}_i$ over $L$ steps, 
where $L$ is the number of Langevin dynamics steps. And $r$ is used to control the 
step length of the correction. The correction schedule can be described as:
\begin{equation}
    {\mathbf{z}_i}^j = {\mathbf{z}_i}^{j-1} + \mathbf{\bm{\xi}}\mathbf{s_{\theta^{*}}}({\mathbf{z}_i}^{j-1}, i) + \sqrt{2\mathbf{\bm{\xi}}}\mathbf{\bm{\varepsilon}},
\end{equation}
where $j$ is from 1 to $L$. Once this round of corrections is finished, we set $\mathbf{z}_i = {\mathbf{z}_i}^{L}$.
The whole sampling algorithm is shown in Algorithm \ref{alg:sampling}.
\begin{algorithm}[htbp]
    \caption{Sampling Algorithm}
    \label{alg:sampling}
    \begin{algorithmic}[1]
        \STATE \textbf{Input:} $\mathbf{z}_{N_{\text{snr}}} \leftarrow \mathbf{\tilde{z}}$, SNR, $N_{\text{snr}}$, $L$, $r$
        \FOR{$i = N_{\text{snr}} - 1$ to $0$}
            \STATE $\mathbf{z}_i' \gets \mathbf{z}_{i+1} + (\sigma_{i+1}^2 - \sigma_i^2) s_{\theta^{*}}(\mathbf{z}_{i+1}, \sigma_{i+1})$
            \STATE $\bm{\varepsilon} \sim \mathcal{CN}(0, \mathbf{I}),$ $\mathbf{g} \leftarrow s_{\theta^{*}}(\mathbf{z}_{i+1}, \sigma_{i+1})$
            \STATE $\bm{\xi}_i \leftarrow 2 \left( r \frac{\|\bm{\varepsilon}\|_2}{\|\mathbf{g}\|_2} \right)^2$
            \STATE $\mathbf{z}_i \gets \mathbf{z}_i' + \sqrt{\sigma_{i+1}^2 - \sigma_i^2} \, \bm{\varepsilon}$
            \FOR{$j = 1$ to $L$}
                \STATE $\bm{\varepsilon} \sim \mathcal{CN}(0, \mathbf{I})$
                \STATE $\mathbf{z}_i \gets \mathbf{z}_i + \bm{\xi}_i s_{\theta}(\mathbf{z}_i, \sigma_i) + \sqrt{2 \bm{\xi}_i} \, \bm{\varepsilon}$
            \ENDFOR
        \ENDFOR
        \STATE $\mathbf{\hat{z}} \leftarrow \mathbf{z}_0$
        \RETURN $\mathbf{\hat{z}}$
    \end{algorithmic}
\end{algorithm}
\subsection{Stage 2 : Joint Training of SCDM and Semantic Decoder for Digital Semantic Communications}
After the SCDM was trained, we can integrate it into the DSC framework to achive an 
approximate reconstruction of $\mathbf{z}$, which is $\mathbf{\hat{z}}$. But the distribution of $\mathbf{\hat{z}}$
is still slightly different from the real distribution of $\mathbf{z}$. So we need jointly retrain the semantic decoder and SCDM
to achieve a better reconstruction of $\mathbf{x}$. The reconstruction loss fuction can be written as: 
\begin{equation}
    \begin{split}
        \label{loss}
        \mathcal{L}(\varphi) = & \mathbb{E}_{\mathbf{\hat{\mathbf{z}}}\sim \mathbf{p}_{\hat{\mathbf{z}}|\mathbf{x}}} \| \mathbf{x} - \hat{\mathbf{x}} \|_2^2. 
    \end{split}
\end{equation} 
 
The full training pipeline of the SCDM-DSC framework is summarized in Algorithm \ref{alg:SCDM-DSC}.
\begin{figure}[htbp]
    \centering
    \includegraphics[width=\linewidth]{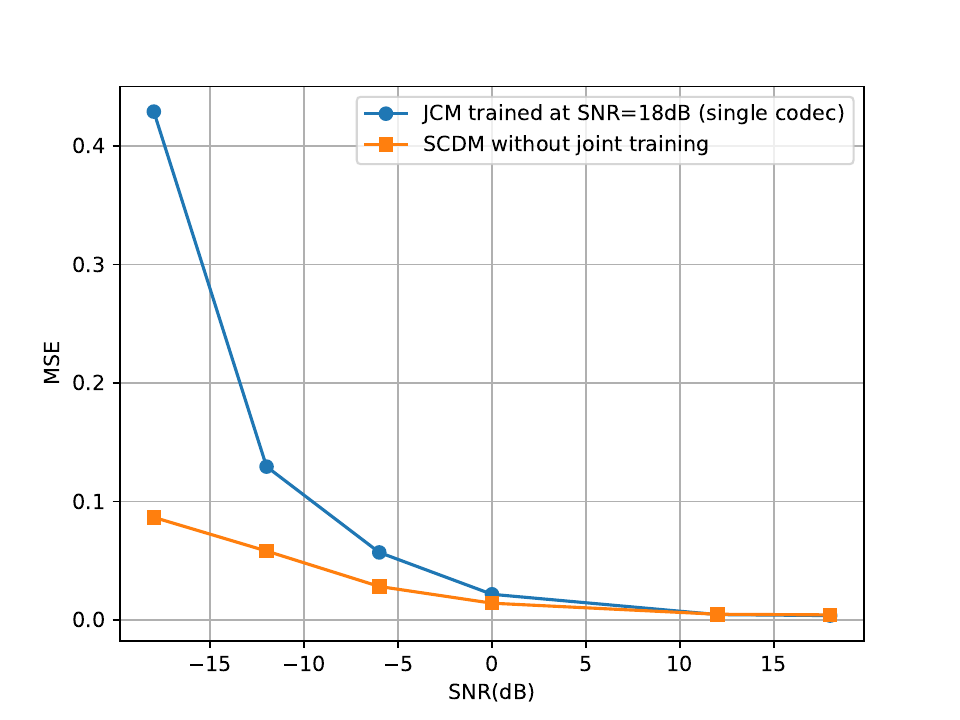}  
    \caption{Comparison of the average MSE of latent semantic code between our proposed SCDM and the baseline method JCM in 64-QAM modulation. The SNR range is set from -18 dB to 18 dB.}
    \label{fig:2}
\end{figure}
\begin{algorithm}[htbp]
    \caption{Training algorithm of the SCDM-DSC }
    \label{alg:SCDM-DSC}
    \begin{algorithmic}[1]
        \STATE \textbf{Input:} $\mathcal{X}$, $N$, $\mathcal{E}$, $\mathcal{D}$, $\phi$, $\varphi$, $\theta$
        \STATE \textbf{Stage 1:}
            \STATE Train SCDM with Algorithm~\ref{alg:scdm_training} 
        \STATE \textbf{Stage 2:}
        \STATE \textbf{repeat}
            \STATE $\mathbf{x}$ $\sim$ $\mathcal{X}$, $N_{snr}$ $\sim$ $\text{Uniform}(\{1, \ldots, N\})$
            \STATE $\mathbf{z}_{0}$ = $\mathcal{E}(\phi)$ ($\mathbf{x}$),  $\mathbf{\bm{\varepsilon}}$ $\sim$ $\mathcal{CN}(0, \mathbf{I})$, 
            compute $\mathbf{z}_{N_{snr}}$ via \eqref{eq:5}
            \STATE $\mathbf{\tilde{z}} \leftarrow {\mathbf{z}_N}_{\text{snr}}$; denoise $\mathbf{\tilde{z}}$ using Algorithm~\ref{alg:sampling} to obtain $\mathbf{\hat{z}}$.
            \STATE Compute $\mathcal{L}(\varphi)$ via \eqref{loss} and update $\varphi$
        \STATE \textbf{until} Converged
    \end{algorithmic}
\end{algorithm}

\section{Experiments}
This section presents extensive experiment results to validate the advantages of the proposed SCDM-DSC framework under various transmission rates, modulation orders, 
and channel conditions. Our experiments were conducted on two NVIDIA GeForce RTX 4090 GPUs, each with 24GB of memory.

\begin{figure*}[htbp] 
    \centering
    \begin{subfigure}[b]{0.45\linewidth}
        \centering
        \includegraphics[width=\linewidth]{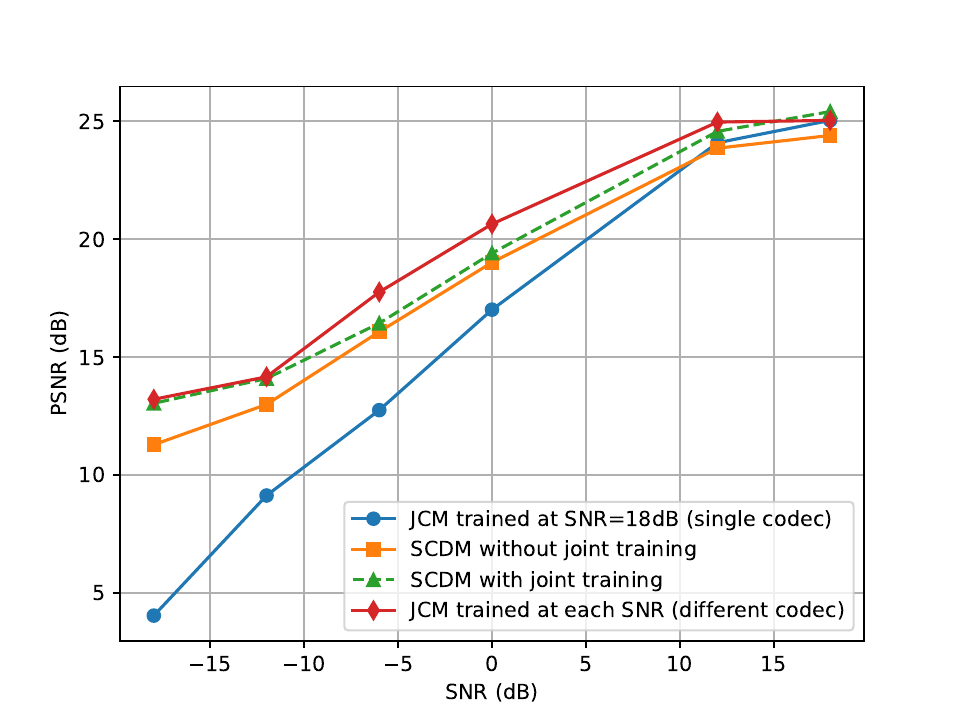}
        \caption{PSNR}
        \label{fig:4a}
    \end{subfigure}
    \hfill
    \begin{subfigure}[b]{0.45\linewidth}
        \centering 
        \includegraphics[width=\linewidth]{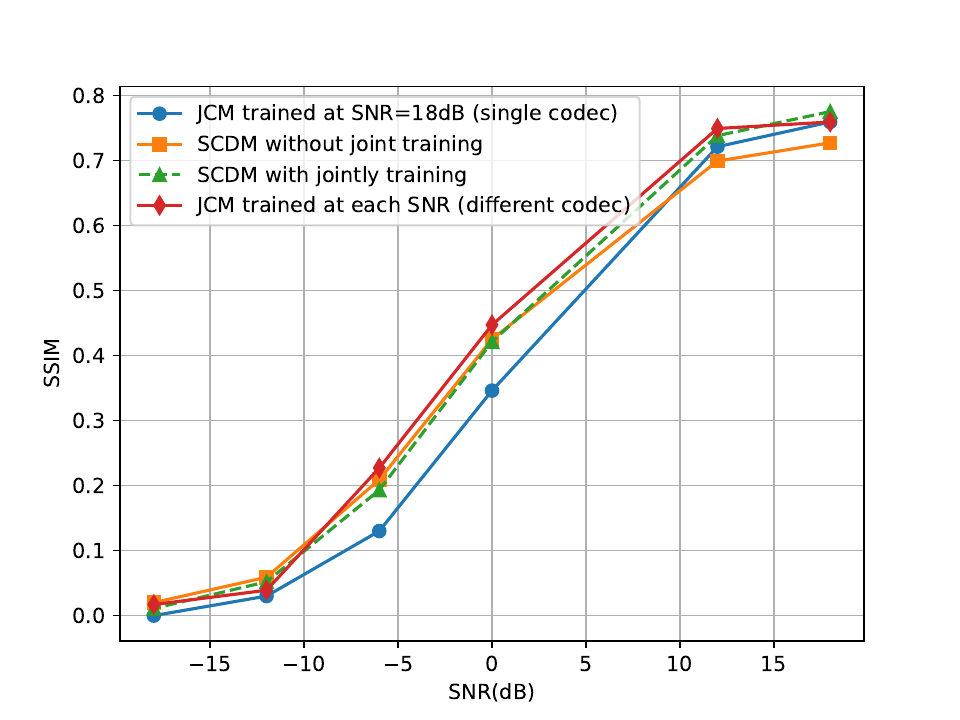}
        \caption{SSIM}
        \label{fig:4b}
    \end{subfigure}
    \caption{Comparison of PSNR and SSIM vs. SNR for SCDM and JCM in 64-QAM modulation.}
    \label{fig:psnr_ssim} 
\end{figure*}

\subsection{Experimental Setup}

\textit{a) Dataset:} Our experiments are conducted on CIFAR-10 \cite{krizhevsky2009learning}, a widely used dataset for image classification. 
CIFAR-10 consists of 60,000 32$\times$32 color image. We use the standard training and testing split method, 
with 50,000 images for training and 10,000 images for testing.

\textit{b) Models:} For DSC system, we use the same Variational Autoencoder (VAE) neural network architecture as presented in \cite{bo2024joint} as our codec. 
In this setup, the input image resolution is 32$\times$32, and the semantic latent code dimension is 2$\times$$n$ in BPSK modulation, and $2\sqrt{M}$$\times$ in M-QAM modulation,
where $n$ represents the length of the constellation symbol sequence, and $M$ represents the modulation order. 

In terms of SCDM, the backbone of the VE SDE is a U-Net \cite{ronneberger2015u} that consists of 3 down-sampling 
blocks, 1 bottleneck block, and 3 up-sampling blocks. Each down-sampling or up-sampling block is a hybrid of a 
ResNet block and a Transformer block. The bottleneck block consists of two ResNet blocks with a Transformer block 
in the middle. We set the diffusion step \(N = 64\), the Langevin dynamics correction step \(L = 2\), and the step 
length \(r = 0.16\).

\textit{c) Training:} We leverage a pretrained VAE model to generate the semantic latent code, which subsequently serves as the input for our second-stage VE SDE model.
We employ the Adam optimizer with a learning rate of 0.0001 throughout the training process. 
  
\subsection{Simulations and Discussions}
We compare our SCDM-DSC framework with the state-of-the-art methods under various channel conditions. We choose JCM as our baseline method.
Our experiments are conducted with the number of channel uses $n$ = 128 and the order of digital modulation $M$ = 64. We also vary the SNR from -18 dB to 18 dB.
We adopt mean sequare error (MSE), average peak signal-tonoise ratio (PSNR), and structural similarity index (SSIM) as our performance metrics.
We also discuss the model size and the disk storage consumption of the SCDM and JCM models.

Fig.~\ref{fig:2} compares the MSE between the latent semantic code before transmission over the digital wireless channel and before being sent to the semantic 
decoder in our 64-QAM simulation for both the proposed SCDM and JCM methods. The codec used for training our SCDM is a 
pretrained JCM model at an 18 dB SNR. Notably, we achieve an improvement of approximately 0.34 in MSE at SNR = -18 
dB. This implies that, under our SCDM-DSC framework, the semantic decoder does not need to contend with the high noise 
variance of the received symbols as in JCM, thereby enhancing the robustness of the entire system under low SNR 
conditions.
 
As shown in Fig.~\ref{fig:psnr_ssim}, we analyze the average PSNR and SSIM of the reconstructed images 
achieved by our SCDM approach and the baseline JCM model across different SNR levels in a 64-QAM modulation 
scheme. From Fig.~\ref{fig:4a}, we have the following observations.

\begin{enumerate} 
    \item In low SNR regions (e.g., SNR $\leq 10$ dB), our SCDM model demonstrates a clear advantage over a 
    single JCM codec trained at 18 dB SNR, even without joint training. For example, we observe a PSNR gain of 
    up to 3.3 dB at an SNR of -6 dB and 7.2 dB at an SNR of -18 dB. This result suggests that the SCDM module 
    effectively mitigates noise distortion in the received digital semantic symbol at low SNR levels, providing 
    a smaller noise variance for the decoder to reconstruct the semantic information of the source data. 
    
    \item Across all SNR levels, SCDM achieves comparable PSNR performance to JCM codecs trained at specific 
    SNR levels, with only a slight average difference of 1.36 dB. Notably, we do not need to train a separate 
    codec for each SNR level. This design reduces the memory and computational resources required for model 
    storage and processing, making the SCDM method more efficient and practical for real-world applications.
\end{enumerate}

We also compare the SSIM performance of the SCDM and JCM models in Fig.~\ref{fig:4b}. From the results, we have the following observations.

\begin{enumerate}
    \item With joint training of SCDM and the decoder, we observe that the overall SSIM performance closely 
    approaches that of JCM codecs trained at specific SNR levels, even surpassing them when the SNR 
    exceeds 15 dB. This demonstrates the robustness of SCDM under varying channel conditions.
    
    \item Joint training for SCDM offers similar SSIM performance to non-joint training at low SNR 
    (e.g., 0 dB), with minor differences. However, at higher SNR levels (e.g., 12 dB and 18 dB), joint 
    training shows clear improvements in SSIM, enhancing image quality as SNR increases.
\end{enumerate}

\begin{figure}[htbp]
    \centering
    \includegraphics[width=\linewidth]{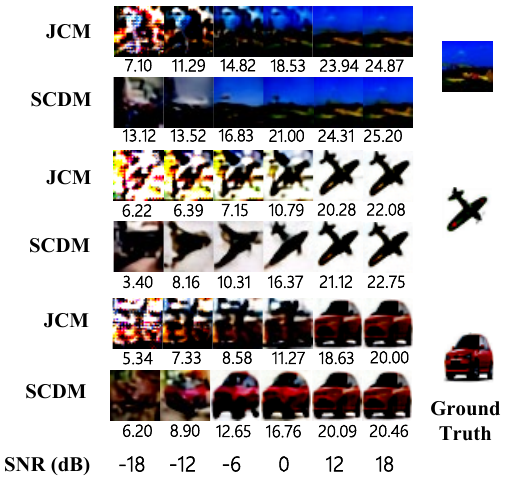}  
    \caption{Visual comparison of the reconstructed image between JCM and SCDM in 64-QAM modulation. The performance metric labeled below each image is PSNR.}
    \label{fig:5}
\end{figure} 

\begin{table}[htbp]
    \centering
    \renewcommand{\arraystretch}{1.5} 
    \setlength{\tabcolsep}{20pt} 
    \begin{tabular}{|c|c|c|}
        \hline
        \textbf{Model name} & \textbf{JCM} & \textbf{SCDM} \\
        \hline
        \textbf{Model size (MB)} & 79.94 & 656.84 \\ 
        \hline
    \end{tabular}
    \caption{Model size comparison between JCM and SCDM.}
    \label{tab:1}
\end{table}

From Fig.~\ref{fig:5}, it can be observed that SCDM and JCM exhibit similar performance when the SNR $\leq 12$ dB. 
However, for SNR values within \([-12, 0]\) dB, the JCM model begins to suffer from severe noise in image reconstruction, 
with the images becoming increasingly distorted as the SNR decreases. In contrast, while the reconstruction quality of SCDM 
also degrades with lower SNR, the semantic information of the images remains largely preserved. For example, 
at $\text{SNR} = -6$ dB, the SCDM model is still able to retain key semantic features of a red car, such as its shape and 
color. In comparison, the image reconstructed by JCM at the same SNR level is visually challenging to recognize as a car, 
indicating a significant loss of semantic integrity.

As shown in Table~\ref{tab:1}, the JCM model requires individual training for each specific SNR value. Within the SNR 
range of \([-18, 18]\) dB, assuming each JCM model covers a 0.5 dB denoising range, a total of 72 JCM models would be 
required, resulting in 5,755.68 MB of storage space. In contrast, our SCDM-DSC framework, consisting of the SCDM and JCM 
components as a codec, requires only 736.78 MB of storage, achieving a storage reduction by a factor of approximately 7.8 
compared to the JCM scheme.

\section{CONCLUSION}
In this paper, we presented a SCDM module for DSC systems, which uses a score-based diffusion model 
to effectively denoise received symbols by aligning with the inherent noise characteristics of digital 
channels. The independent, plug-and-play design of SCDM allows it to function seamlessly with 
existing DSC systems without requiring additional integration into the training process. Moreover, 
its unified architecture accommodates a broad range of SNR levels, significantly reducing storage needs and 
enhancing deployment efficiency. Simulation results demonstrate that SCDM consistently outperforms the baseline 
JCM (single codec case) across PSNR, SSIM, and MSE metrics, particularly under low SNR conditions, 
highlighting its robustness and adaptability to varying channel conditions. 
This advancement promotes the development of more scalable and robust semantic communication systems for future 
6G networks.

\section{ACKNOWLEDGMENT}
This work was supported in part by National Natural Science Foundation of China (NSFC) under Grants 62301471, 62222111, 62293482, 62125108; 
in part by the National Science and Technology Major Project - Mobile Information Networks 
under Grant No.2024ZD1300700.

\bibliography{references}

\end{document}